\documentclass[aps,twocolumn,floatfix,prl,letterpaper,superscriptaddress]{revtex4-1}
\usepackage{amsmath}
\usepackage{hyperref}
\usepackage{amssymb}
\usepackage{graphicx}
\usepackage{epsf}
\usepackage{epstopdf}
\usepackage[dvipsnames]{xcolor}

\usepackage{dsfont}
\usepackage{float}
\usepackage{scrextend}

\allowdisplaybreaks

\newcommand{\beq}{\begin{equation}}
\newcommand{\eeq}{\end{equation}}

\newcommand{\bpm}{\begin{pmatrix}}
\newcommand{\epm}{\end{pmatrix}}

\newcommand{\ba}{\begin{array}{l}}
\newcommand{\ea}{\end{array}}

\newcommand{\R}{\mathbb{R}}

\newcommand{\B}[1]{\boldsymbol{#1}}

\begin{document}

\title{Topology, geometry and mechanics of strongly stretched and twisted filaments}
\date{\today}

\author{Nicholas Charles}
\affiliation{Paulson School of Engineering and Applied Sciences,
Harvard University, Cambridge, MA 02138}
\author{Mattia Gazzola}
\affiliation{Department of Mechanical Science and Engineering, University of Illinois at Urbana-Champaign, Urbana, IL 61801, USA}
%\affiliation{National Center for Supercomputing Applications, Urbana, IL 61801, USA}
\author{L. Mahadevan}
\affiliation{Paulson School of Engineering and Applied Sciences, Department of Physics, Department of Organismic and Evolutionary Biology, Harvard University, Cambridge, MA 02138}
\email{lmahadev@g.harvard.edu}

\begin{abstract}
Soft elastic filaments that can be stretched, bent and twisted exhibit a range of topologically and geometrically complex morphologies that include plectonemes, solenoids, knot-like and braid-like structures. We combine numerical simulations of soft elastic filaments that account for geometric nonlinearities and self-contact to map out these structures in a  phase diagram that is a function of extension and twist density, consistent with previous experimental observations.  By using ideas from computational topology, we also track the interconversion of link, twist and writhe in these geometrically complex physical structures. This allows us to explain recent experiments on fiber-based artificial muscles that use the conversion of writhe to extension or contraction, exposing the connection between topology, geometry and mechanics in an everyday practical setting.  

\end{abstract}

\maketitle

The bending and twisting elastic response of soft filamentous objects is a consequence of the geometric separation of scales. This realization is at the heart of the classical Kirchhoff-Love theory \cite{Kirchhoff1859, Love1892} which considers inextensible, unshearable filaments and has spawned a substantial literature \cite{Antman2004, OReilly2017}. When such filaments are twisted strongly, they deform into plectonemic structures that consist of self-braided segments and have been seen in filaments across scales, from DNA to metal wires \cite{Schlick2000, Coyne1990}. The transition between the straight and plectonemic structures in inextensible filaments has been explored extensively in both a deterministic and a stochastic setting \cite{Stump1998,Thompson2000, Olson1991, Marko2012}, and continues to be a topic of interest.  However filaments made of soft elastomeric materials are extensible and shearable, and their study is interesting for a range of applications such as biological tissue mechanics, soft robotics etc. Amongst the simplest behaviors that harnesses these modes of deformation is the controlled transition between straight filaments and tightly coiled helical shapes (solenoids), originally observed in textiles \cite{Hearle1972}, quantified experimentally in elastomers \cite{Ghatak2005}, and then rediscovered in the context of heat-driven artificial muscles \cite{Baughman2016}. These energy harvesting devices rely on the conversion of twist and bend into extension \cite{Haines2017, Kim2018, Atikah2017}, as solenoids untwist and stretch. Here we consider the interplay between topology, geometry and mechanics in strongly stretched and twisted filaments to explore the range of morphologies seen and their functional consequences. 
%Figure 1abc
\begin{figure}
\begin{center}
\includegraphics[width = 0.5 \textwidth]{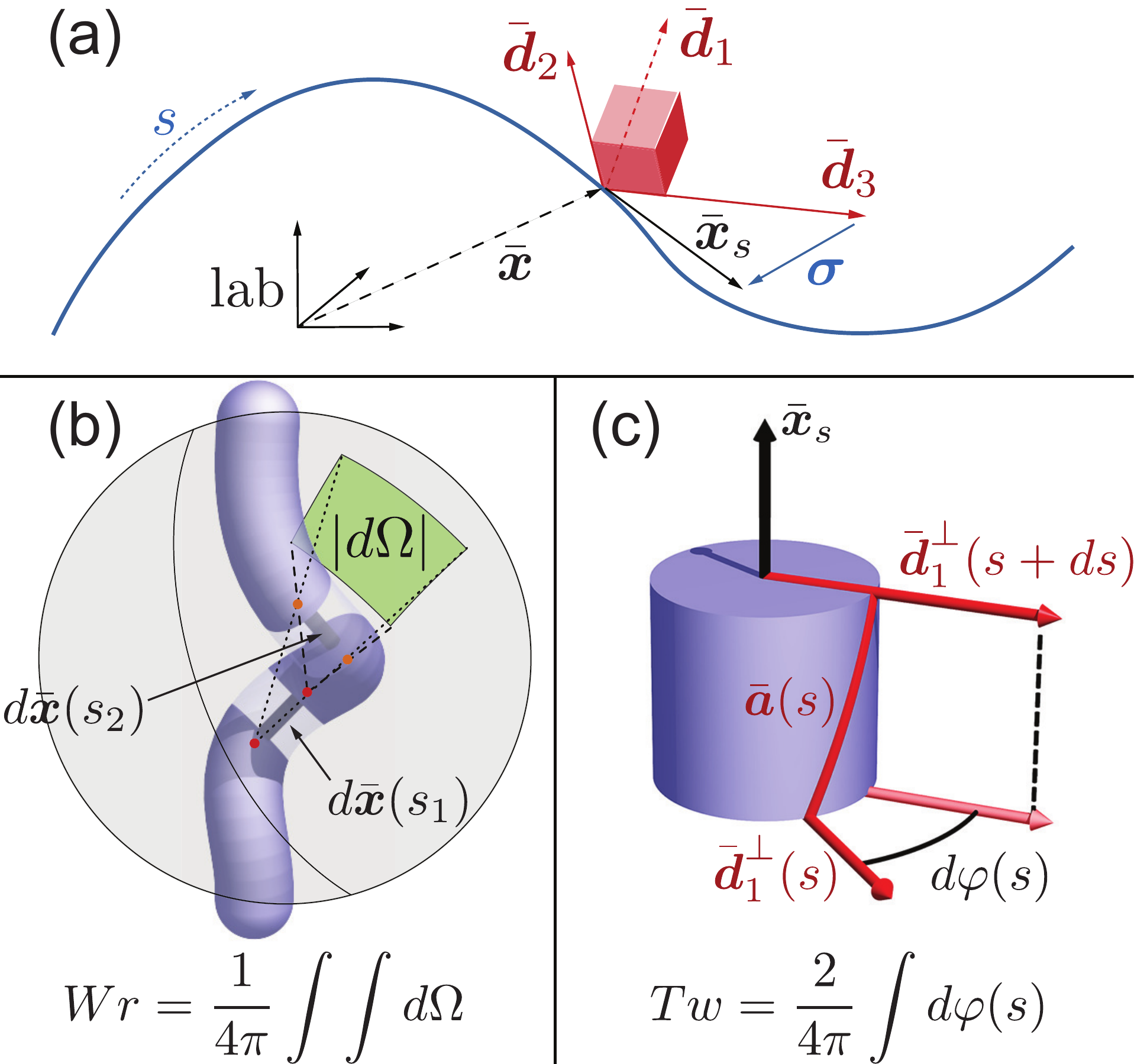}
\caption{Geometry and topology of soft extensible filaments. \textbf{(a)} The filament centerline $\boldsymbol{\bar{x}}(s,t)$ and local orthogonal frame  $\{\B{\bar{d}}_1, \B{\bar{d}}_2, \B{\bar{d}}_3 \}$. Shear and extension are defined by the vector $\boldsymbol{\sigma} = \boldsymbol{Q} \boldsymbol{\bar{x}}_s - \boldsymbol{d}_3$, while curvature and twist are defined by the vector ${\bf k} = {\rm vec} ({\bf Q' Q^T})$. \textbf{(b)} Writhe (\textit{Wr}) equals the centerline's average oriented self-crossing number, computed in terms of the integral of the solid angle $d \Omega$ determined by the infinitesimal centerline segments $\B{\bar{x}}(s_1)$ and $\B{\bar{x}}(s_2)$ (left-handed intersections are negative). \textbf{(c)} Twist (\textit{Tw}) is the integral of the infinitesimal rotations $d\varphi$ of the auxiliary curve $\B{\bar{a}}$ around $\B{\bar{x}}_s$. Here the vector $\B{\bar{a}}$ traced out by $\B{\bar{d}}_1^\bot$ (i.e., the projection of $\B{\bar{d}}_1$ onto the normal-binormal plane) is shown in red while the curve associated with $-\B{\bar{d}}_1$ is shown in yellow (see Fig.~\ref{slices}). For a closed curve $Lk = Tw + Wr$, where \(Lk\) (link) is the average oriented crossing number of $\B{\bar{x}}(s)$ with $\B{\bar{a}}(s)$.}
\label{framework}
\end{center}
\vspace{-22pt}
\end{figure}

We describe a filament by a centerline position vector $\B{\bar{x}}(s, t) \in \R^3$ ($s \in [0, L_0]$ is the material coordinate of  the rod of rest length $L_0$ at time $t$), while the orientation of its cross-section is defined by an initially orthonormal triad associated with the director vectors $\B{\bar{d}}_i(s,t), i = 1,2,3 $, where $\B{\bar{d}}_3(s,t)$ is normal to the material cross-section of the filament. Then, the transformation of the body-fixed frame to the lab-fixed frame can be written in terms of the rotation matrix $\B{Q}(s, t)  = \{ \B{\bar{d}}_1, \B{\bar{d}}_2, \B{\bar{d}}_3 \}^{-1}$ (see Fig.~\ref{framework}a).  

In general, the centerline tangent \(\partial_s \B{\bar{x}} = \B{\bar{x}}_s\) does not point along the normal to the cross-section \(\B{\bar{d}}_3(s,t)\). The deviation between these vectors characterizes local extension and shear \(\B{\sigma} = \B{Q} (\B{\bar{x}}_s - \B{\bar{d}}_3) = \B{Q}\B{\bar{x}}_s - \B{d}_3\) (Fig.~\ref{framework}a), and is the basis of the  Cosserat rod theory \cite{Antman2004}, that allows us to include all six modes of deformation at every cross-section (mathematically, this is associated with the dynamics on the full Euclidean group $SE(3)$). The restriction to the Kirchhoff theory corresponds to the case \(\B \sigma = 0\), i.e. the normal to the cross-section is also the tangent to the centerline,  with \({\bar{x}}_s - \B{d}_3 = 0\).   

Since many soft materials are close to being incompressible (i.e. the shear modulus is much smaller than the bulk modulus), we assume the filament material to be incompressible. Then, if \( e = |\B{\bar{x}}_s| \) is the elongation factor and \( A \) is cross-sectional area, \( Ae \) is constant. This allows us to use a simple materially linear constitutive law that is a reasonable approximation to both neo-Hookean and Mooney-Rivlin materials  (see \cite{Gazzola2016} and SI for validation and \cite{Spillmann2007, Audoly2013, Bergou2008} for alternative approaches).  

Then, we may write the linear and angular momentum balance equations as \cite{Antman2004, OReilly2017, Gazzola2016}
\begin{align*}
\rho A \cdot \partial_t^2 \B{\bar{x}} = & \partial_s \left( \frac{\B{Q}^T \B{S} \B{\sigma}}{e} \right) + e\B{\bar{f}}\\
\frac{\rho \B{I}}{e} \cdot \partial_t \B{\omega} = & \partial_s \left( \frac{\B{B} \B{k}}{e^3} \right) + \frac{\B{k} \times \B{B} \B{k}}{e^3} + \left( \B{Q}\frac{\B{\bar{x}}_s}{e} \times \B{S} \B{\sigma} \right) \\
& + \left( \rho \B{I} \cdot \frac{\B{\omega}}{e} \right) \times \B{\omega} + \frac{\rho \B{I} \B{\omega}}{e^2} \cdot \partial_t e + e\B{c}
\end{align*}
%\begin{align*}
%0 = & \partial_s \left( \frac{\B{Q}^T \B{S} \B{\sigma}}{e} \right) + e\B{\bar{f}}\\
%0 = & \partial_s \left( \frac{\B{B} \B{k}}{e^3} \right) + \frac{\B{k} \times \B{B} \B{k}}{e^3} + \left( \B{Q}\frac{\B{\bar{x}}_s}{e} \times \B{S} \B{\sigma} \right)
%\end{align*}
where $\rho$ is the material density, $\B{\bar{\omega}}={\rm vec} (\partial_t \B{Q}^T \B{Q})$ is the local angular velocity, $\B{\bar{k}} = {\rm vec} (\partial_s \B{Q}^T \B{Q})$ is the local strain vector (of curvatures and twist), $\B{S}$ is the matrix of shearing and extensional rigidities, $\B{B}$ is the matrix of bending and twisting rigidities,  and $\B{f}, \B{c}$ are the  body force density and external couple density (see SI or \cite{Gazzola2016} for details).
%Figure 2ab. 
\begin{figure}
\begin{center}
\includegraphics[width = 0.5 \textwidth]{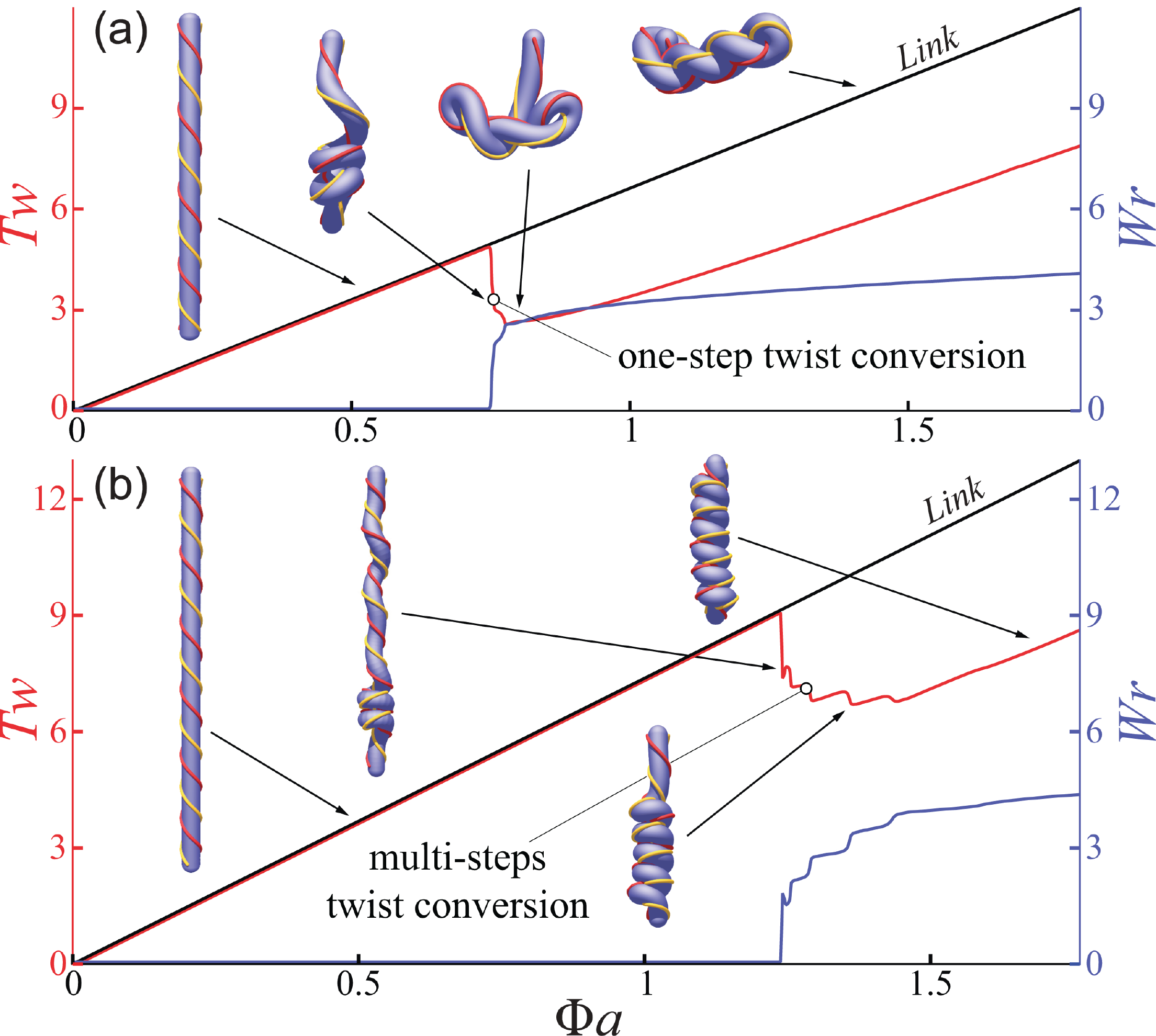}\\
\caption{Variation of the link, twist and writhe as a function of the dimensionless twist density $\Phi a$. \textbf{(a)} To replicate the  experimental observations in \cite{Ghatak2005}, we use a constant vertical load $F$$\approx$$25$$F_C$ to produce a plectoneme ($F_C$$=$$\pi^2 E I/L_0^2$ is  buckling force for an inextensible rod--see Movie S1). \textbf{(b)} We repeat the simulation with $F$$\approx$90$F_C$, stretching the filament by $L/L_0$$\approx$$1.16$. Increased stretching leads to an overall similar conversion of twist to writhe  leading to tightly packed solenoidal structures (See Movie S2 and SI for plots of filament energy). Simulation settings (SI): length $L_0$$=$$1$ m, $a$$=$$0.025L_0$, Young's modulus $E$=$1$ MPa, shear modulus $G$=$2E/3$, $\boldsymbol{S}$=$\text{diag}(4 G A/3, 4 G A/3, E A)$ N, $\boldsymbol{B}$=$\text{diag}(EI_1, EI_2, GI_3)$ N$\text{m}^2$.}
\label{slices}
\end{center}
\vspace{-22pt}
\end{figure}

To  follow the geometrically nonlinear deformations of the filament described by the equations above, we employ a recent simulation framework \cite{Gazzola2016}, wherein the filament is discretized in a set of $n+1$ vertices $\{\B{\bar{x}}_i\}_{i=0}^n$ connected by edges $\B{\bar{e}}^i = \B{\bar{x}}_{i+1} - \B{\bar{x}}_i$, and a set of $n$ frames $\{ \B{Q}^i\}_{i=0}^{n-1}$. The resulting discretized system of equations is integrated using an overdamped second order scheme, reducing the dynamical simulation to a quasi-static process, and accounting for self-contact forces (SI and \cite{Gazzola2016} for details) while ignoring friction \footnote{While our algorithms can account for friction \cite{Gazzola2016},   the inclusion of this effect does not qualitatively affect our results and so we ignore it herein.}

%Figure 3
\begin{figure}
\begin{center}
\includegraphics[width = 0.5\textwidth]{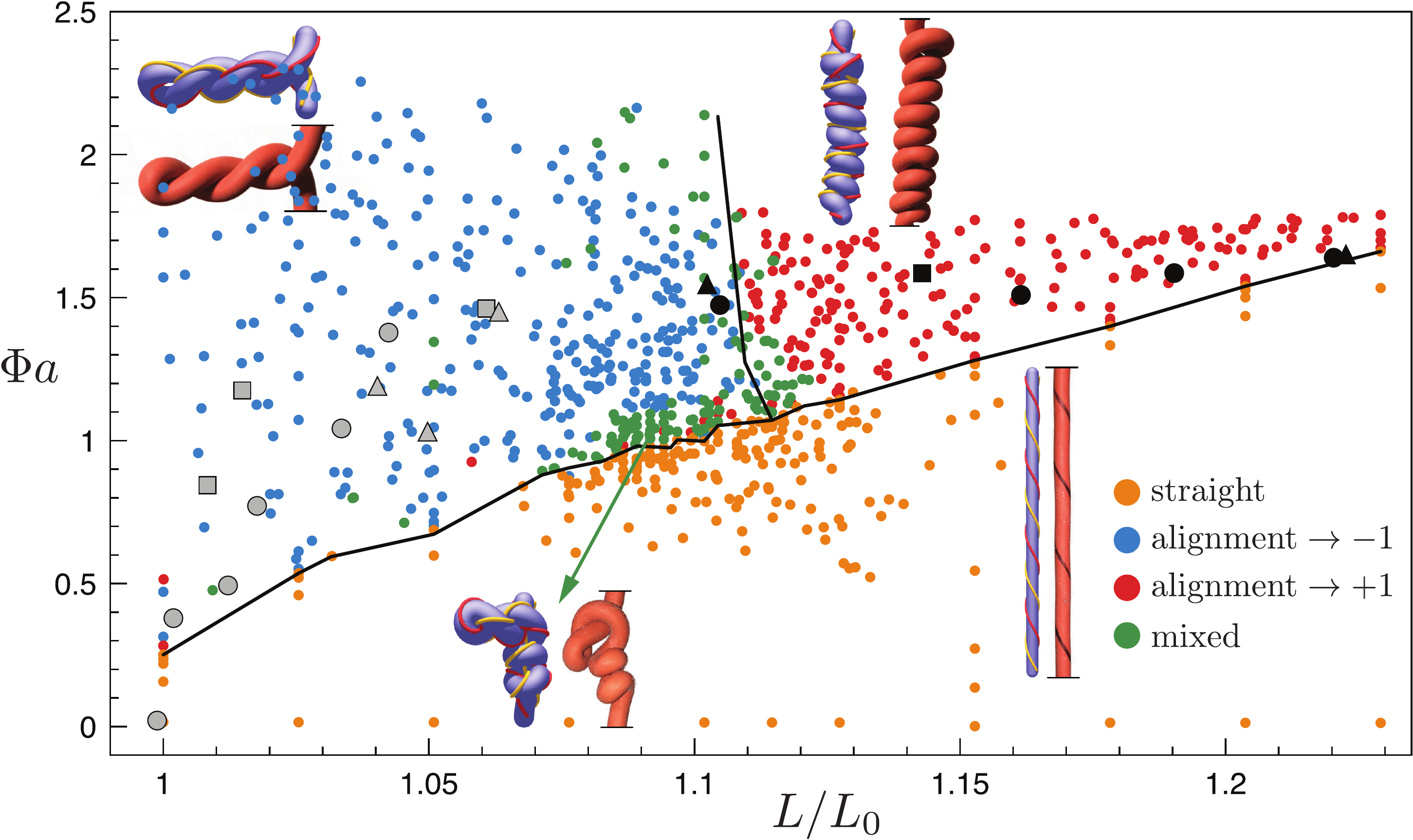}
\caption {Morphological phase space. We simulate a filament prestretched to $L/L_0$ by a constant axial load and twisted by an angle $\Phi a$, as in Fig.~\ref{slices}. By computing centerline relative alignment in neighboring loops, we find four phases: straight, plectoneme, solenoid and plectoneme-solenoid combinations. Plectoneme alignment \(\approx\)\(-1\), solenoid alignment \(\approx\)\(1\) and transition configuration alignments approach 0 (dark green). For $L/L_0 \gtrsim 1.1$ solenoids are preferred. We expect $\Phi_{\text{critical}}$ to scale linearly with $L/L_0$ at high extension, in agreement with this plot. Our results agree qualitatively with experiments \cite{Ghatak2005} (shown in black dots, see SI for details). Hollow symbols denote plectoneme transitions while solid points denote solenoid transitions; different shapes correspond to different filament parameters (SI).  Simulation settings (SI): \(L_0\)\(=\)\(1\) m, \(a\)\(=\)\(0.025L_0\), \(E\)\(=\)\(1\) MPa, \(G\)\(=\)\(2E/3\), \(\boldsymbol{S}\)\(=\)\(\text{diag}(4 G A/3, 4 G A/3, E A)\) N, \(\boldsymbol{B}\)\(=\)\(\text{diag}(EI_1, EI_2, GI_3)\) N\(\text{m}^2\).}
\label{phase}
\end{center}
\vspace{-22pt}
\end{figure}

To track the knot-like structures that form when the stretched and twisted filament can contact itself, we take advantage of the CFW theorem \cite{Fuller1978, Calugareanu1959}: \textit{Link} (\textit{Lk}) $=$ \textit{Twist} (\textit{Tw}) $+$ \textit{Writhe} (\textit{Wr})  Here, link is the oriented crossing number (or Gauss linking integral) of the centerline and auxiliary curve $\B{\bar{a}}(s)$ (Fig.~\ref{framework}) averaged over all projection directions \cite{Ricca2010}, writhe is the link of the centerline with itself \cite{Fuller1971}, and twist denotes the local rotation of the auxiliary curve about the centerline. In a discrete setting, we compute writhe, link and twist of the filament modeled as an open ribbon following \cite{Klenin2000}, as illustrated in Fig.~\ref{framework} (see the SI for details).  

When inextensible filaments are stretched and twisted, a range of localized and self contacting structures arise, and have been well studied in both a deterministic and stochastic setting \cite{Thompson1998, Goriely1998, Olson1991, Nelson1997, Marko2012, Gerbode2012}. For highly stretched and twisted filaments, the phase space of possibilities is much richer, and in particular a new morphological phase associated with tightly coiled helices (solenoids) appears \cite{Hearle1972,Ghatak2005}. To characterize these morphologies, we first twist a filament clamped at one end and prestretched by a constant axial load \(\approx\)25 times the critical compressive buckling force of a corresponding inextensible filament $F_C$=$\left( \pi^2 E I \right)/L_0^2$. In Fig.~\ref{slices}a, we show that when  a critical dimensionless twist density $\Phi a$ is reached, the filament becomes unstable to bending, leading to the formation of a plectoneme, converting twist to writhe; occasionally the plectoneme can partially untie itself by slipping a loop over an endpoint, allowing link to escape the system (Fig.~S3). In Fig.~\ref{slices}b, we repeat the simulation but quadruple the stretching strain and see that at a critical value of $\Phi a$, the filament again becomes unstable to bending, but now leads to a qualitatively different equilibrium configuration: a tightly coiled helical solenoid.  We note that  substantial prestretch is the crucial prerequisite for solenoid formation, while shearing is found to be unimportant (see SI for details). While both plectonemes and solenoids convert twist to writhe in steps, they are otherwise quite different. Plectonemes lead to braids made of oppositely chiral helices, while solenoids lead to a single compact helix. Furthermore, a plectoneme loop converts much more twist to writhe than a solenoid does as it coils up (Fig.~\ref{slices}). However the tightly-coiled nature of the solenoidal coil makes it more stable under stretching.  

% Figure 4abc 
\begin{figure*}
\begin{center}
\includegraphics[width=\textwidth]{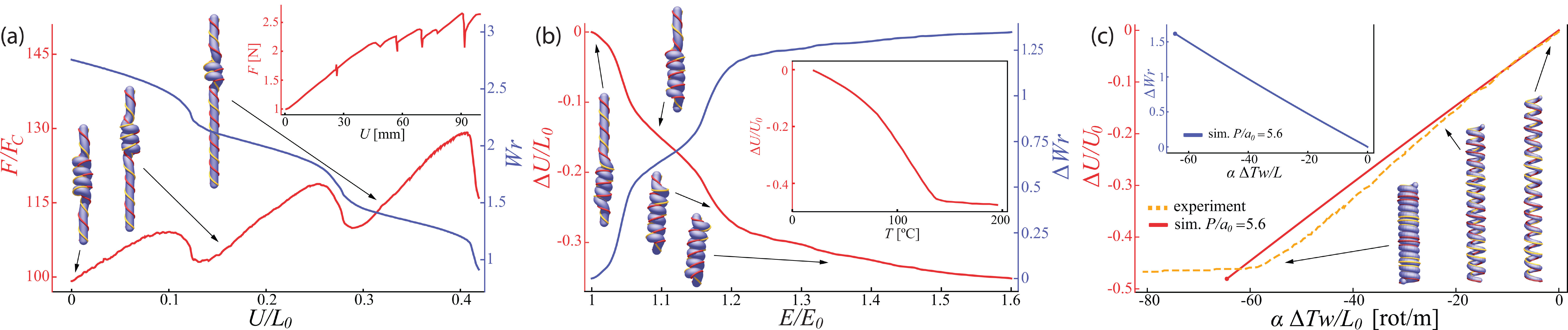}
\caption{Actuation of fiber-based artificial muscles that use the straight-solenoid transition. \textbf{(a)} Passive extension via solenoid loss. \textbf{(a)} We displace the unclamped end $\B{\bar{x}}_n$ of a solenoid formed as in Fig.~\ref{phase} with a  load $\approx$99$F_C$ load a distance $\Delta U$ in the direction $\B{\bar{x}}_n - \B{\bar{x}}_0$ and plot force on $\B{\bar{x}}_n$, qualitatively reproducing experiments \cite{Ghatak2005} (inset--see Movie S13). Simulation settings (SI): $L_0$=$1$ m, $a$=$0.025L_0$, $E$=$1$ MPa, $G$=$2E/3$, $\B{S}$=$\text{diag}(4GA/3, 4GA/3, EA)$ N, $\B{B}$=$\text{diag}(EI_1, EI_2, GI_3)$ N$\text{m}^2$.\textbf{(b)} Active work done by changing temperature which effectively increases filament rigidity, here simply modeled by increasing the Young's modulus E of the material. This leads to the formation of a solenoidal loop in a stretched twisted filament as in Fig.~\ref{phase} with a load $\approx$116 $F_C$ as $E_0$ increases gradually from 1 MPa, showing displacement $\Delta U$ of $\B{\bar{x}}_n$ and increase in writhe $\Delta Wr$ from initial coil writhe, reproducing experiments \cite{Baughman2016} (inset--see Movie S3).  \textbf{(c)} Contraction of twisted and coiled nylon polymer muscle formed by inserting twist and annealing into helix.  Filament radius doubles from initial radius \( a_0\)\(=\)\(0.01\) m while twist decreases to keep \(a k_3\) constant.   Numerical slope and onset of self-contact (shown as point)  agree closely with experimental results \cite{Baughman2016} (see SI for details). Beyond self-contact, radial growth pushes adjacent loops farther apart leading to helix elongation. Note that \( \Delta Tw + \Delta Wr < 0 \) in the inset; Indeed, link escapes from the free boundary due to revolution of the free filament endpoint around the helix axis, reducing the number of loops in the helix (see Fig.~S7, Movies S4, S5).  Simulation settings (SI):  $L_0$=$1$ m, $a$=$0.025L_0$, $E$=$30$ GPa, $G$=$2E/3$, $\B{S}$=$\text{diag}(4GA/3, 4GA/3, EA)$ N, $\B{B}$=$\text{diag}(EI_1, EI_2, GI_3)$ N$\text{m}^2$. Note that pitch \(P\), \(\alpha\)\(=\)\(100\), number of loops and helix radius determine \( L_0 \).}
\label{actuate}
\end{center}
\vspace{-22pt}
\end{figure*}

We now turn to explain the experimental observations and morphological phase diagram that span the twist density-extensional strain \((\Phi a)\)-\((L/L_0)\) phase space \cite{Ghatak2005}.  Using randomly-sampled twist densities and extensions in this phase space, we classify each resulting configuration on the spectrum from plectoneme to solenoid using the average relative alignment of tangent vectors at filament segments which are adjoining in absolute coordinates but separated in material coordinates, i.e. \(\text{avg}_{i = 1}^n\left(\text{sign}(\B{\bar{e}}^i \cdot \B{\bar{e}}^k)\right)\) where \(k\)\(=\)\(\text{argmin} \left(|\boldsymbol{\bar{x}}_k -\boldsymbol{\bar{x}}_i|\right)\) subject to \(|k-i| \textgreater \frac{5 n a }{L_0}\) and \(|\boldsymbol{\bar{x}}_k - \boldsymbol{\bar{x}}_i| \textless (2 + \epsilon)a\), with \(\epsilon\)\(=\)\(0.2\) (empirically determined to maximize classification accuracy).  Plectoneme loops involve two strands entwined in antiparallel directions (alignment\(\rightarrow\)\(-1\)), while segments of adjacent solenoid loops tend to lie parallel (alignment\(\rightarrow\)\(1\)), and straight segments do not contribute to the average. In Fig.~\ref{phase} we show four qualitatively different filament configurations: rectilinear, plectoneme, solenoid, and a mixed state with features of both plectonemes and solenoids; indeed the distinction between solenoid and plectoneme becomes blurred near the triple point. These simulations agree qualitatively with experimental observations \cite{Ghatak2005}, as illustrated in Fig.~\ref{phase}; the small quantitative discrepancy between experiments and simulations is likely due to our neglect of friction.   {It is worth pointing out that the region of solenoid-plectoneme coexistence can be changed by having an active agent, e.g. a DNA-binding enzyme, capable of either relaxing the internal axial tension and/or inducing excess twist in the filament locally.  This allows for the formation of a plectoneme  in the compressed segment, after which, upon further twisting, a solenoid forms below the lifted point (Fig. S9 and SI), with similarities to loop formation in chromosomes \cite{Fudenberg}. }

Our results also explain earlier observations \cite{Hearle1972} that describe straight-plectoneme-solenoid transitions in terms of varying twist density and correspond to tracing horizontal and diagonal paths through the present extension-twist density phase diagram (see SI).   Indeed, horizontally exiting the solenoid region in Fig.~\ref{phase} to the right by gradually displacing the lower solenoid endpoint away from the top leads to a step-like solenoid \textit{loss} process.  We track the required force and resulting change in writhe (Fig.~\ref{actuate}a).  The solenoid remains mostly coiled, resisting stretching with a linear force-displacement relation, until a critical displacement at which it uncoils by one pitch and the process starts again. This stepwise uncoiling stems from a kinematic competition similar to solenoid formation: stretching the filament increases the energy required to maintain a constant number of coils. The simulated sawtooth force-displacement pattern agrees qualitatively with experiments \cite{Ghatak2005}. 

We now turn to quantitatively explain a series of recent experiments on a new class of artificial muscle fibers \cite{Baughman2016} that exploit temperature to change the geometry and stiffness of the fibers. The applied temperature change causes an increase in the radius and stiffness of a pre-twisted filament and causes it to untwist, producing an equivalent increase in writhe and work against external loads. The topologically-constrained conversion of twist to writhe as the filament changes shape to lower energy is also seen in the formation of solenoids in highly stretched, twisted filaments \cite{Ghatak2005, Hearle1972}. The main difference is that while in solenoid formation that change is induced by increasing twist externally, in artificial muscles it is induced by using a scalar field, temperature that changes the material properties.   In Fig.~\ref{actuate}b, we simulate this by showing the effects of gradually increasing the elastic modulus in a pre-stretched filament that is subject to a constant load and twisted just beyond solenoid formation.  To increase writhe, the solenoidal state progressively invades the straight state, lifting its lower endpoint toward the clamped end, qualitatively reproducing experimental observations \cite{Baughman2016}.  However, unlike in our simulations, the experimental filament is first annealed around a mandrel with space between adjacent loops before contraction, preventing new loop formation but allowing contraction by decreasing loop spacing.  Hence, the naturally-formed simulated solenoid contracts via the characteristic steplike process of new loop formation.  As noted in the setting of a uniform coil \cite{Lamuta2018}, the contraction due to radial expansion can be modeled by taking into account only the effective increase in stiffness of the underlying coil fiber, as here.  

To fully replicate the actuation experiments in \cite{Baughman2016}, we initialize a filament with intrinsic twist and numerically anneal the filament into a uniform coil with space between adjacent loops, replicating the plastic deformation process by which twisted and coiled polymer muscles are formed.  The fibers used in \cite{Baughman2016} expand radially and contract axially when heated; however, as noted quantitatively in \cite{Lamuta2018}, considering radial growth with fixed fiber length is sufficient.  While our model applies to an isotropic filament rather than one made from aligned polymer chains or carbon nanotubes, we can simulate anisotropic expansion-driven untwist by following the mechanical analogy described in \cite{Baughman2016}: imagine winding an inextensible string around a fiber, attaching it on both fiber ends.  To keep the length of the string constant, the fiber would have to untwist to expand. Mathematically, this requires \( a k_3 \) to stay constant.  Hence, we prescribe a radial growth rate and continuously update the intrinsic twist to keep \( a k^{\text{intrinsic}}_3 \) constant \footnote{Note that we update \textit{intrinsic} twist \(k^{\text{intrinsic}}_3\) rather than true twist \( k_3 \), since \( k_3 \) must evolve according to the equations of motion.  Since the filament twisting strain is defined by \( k_3 - k^{\text{intrinsic}}_3 \), the filament deforms to try to make \( k_3 = k_3^{\text{intrinsic}}\).  Note that after the onset of self-contact, radial growth pushes adjacent loops apart, lowering filament writhe and thereby forcing the filament to retwist despite an ever-increasing intrinsic twist.}.  For a homochiral coil the resulting untwist leads to contraction (Fig.~\ref{actuate}), but in a heterochiral coil to elongation (Fig.~S8, Movie S6, S7).  

In Fig.~\ref{actuate}c we show change in \( Wr \) and contraction for a simulated coil with initial inserted twist density of \(2\) rot/m. Our theoretical computation predicts that an initial pitch \( P_0\)\(=\)\(0.056\) m leads to self-contact at a similar \( \Delta Tw/L_0\) to that in \cite{Baughman2016}. We scale simulated filament parameters to increase simulation efficiency and show contraction over equivalently-scaled twist density, denoting the scaling parameter by \( \alpha\)\(=\)\(100\).  Both coils contract at the same scaled rate as experiments until adjacent loops come into contact (see SI for details  of varying \( P/a_0\)).  

All together, our study links topology, geometry and mechanics in the context of the highly non-trivial morphological behavior of soft, strongly stretched, twisted filaments. This allows us to explain old observations in textile mechanics, quantify recent experiments on artificial muscle fibers and sets the stage for the study of complex braided, knotted and twisted filament configurations of soft filaments in a range of new settings.

\textbf{Acknowledgements.} We thank Andrew McCormick for his preliminary contributions to the project, and the  UIUC Blue Waters project (OCI-0725070, ACI-1238993) (MG), the NSF EFRI C3 SoRo \#1830881 (MG), the NSF CAREER \#1846752 (MG), the Harvard MRSEC NSF DMR 14-20570 (LM) and Harvard BioMatter NSF DMR 33985 (LM) grants  for partial financial support.

\end{document}